\documentclass[aps, showpacs, pra,amssymb, amsmath, twocolumn]{revtex4}
\usepackage{amsmath,latexsym,amssymb}
\usepackage{graphicx}
\usepackage{amsmath}
\usepackage{latexsym}
\usepackage{amssymb}
\usepackage{bm}
\usepackage{placeins}
\usepackage{color}
\usepackage{bbm}
\usepackage{multirow}
\usepackage{cancel}
\usepackage{float}
\usepackage{xcolor}

\newcommand{\bra}[1]{\left\langle #1\right|}
\newcommand{\ket}[1]{\left|#1\right\rangle}
\newcommand{\abs}[1]{\left|#1\right|}

\begin{document}

\title{Subpicosecond $X$ rotations of atomic clock states}
\author{Yunheung Song, Han-gyeol Lee, Hyosub Kim, Hanlae Jo, and Jaewook Ahn}
\address{Department of Physics, KAIST, Daejeon 305-701, Korea}
\date{\today}

\begin{abstract} 
We demonstrate subpicosecond-time-scale population transfer between the pair of hyperfine ground states of atomic rubidium using a single laser-pulse. Our scheme utilizes the geometric and dynamic phases induced during Rabi oscillation through the fine-structure excited state in order to construct an $X$ rotation gate for the hyperfine-state qubit system. Experiment performed with a femtosecond laser and cold rubidium atoms,  in a magneto-optical trap, shows over 98\% maximal population transfer between the clock states.
\end{abstract}

\maketitle

\section{introduction}
High-speed operation of qubit logic gates is crucial in dealing with quantum systems of a limited coherence time~\cite{ultrafastiontwo,ultrafastquntumdot,ultrafastion,ultrafastquantumdottwo,geophasegate,ultrafastmatter}. Since the allowable number of operations is the coherence time divided by qubit operation time, shortening the operation time is equally important as increasing the coherence time. 

Atomic hyperfine states in the ground state can have a long coherence time measured up to tens of seconds~\cite{clock_state_coherence_time}; because of this, they are used not only as the clock states in atomic clocks, which keep the most accurate time and frequency standards~\cite{atomicclock}, but also as storage qubits in quantum computation~\cite{QIreview,QCreview}. Atomic qubits have been controlled with direct microwave transitions~\cite{microwave,RB,3D,MI,ionMW} or CW-laser Raman transitions~\cite{Raman,Raman2,micromirror,ionRaman,ionRaman2}. Experiments~\cite{ionMW,ionRaman,ionRaman2,MI} showed below $10^{-4}$ infidelity of single qubit controls, lower than the commonly accepted error threshold for fault-tolerant quantum computing. The clock speed of these gates typically is from kHz to MHz range.  However, the fundamental limit of the speed is given by the hyperfine energy gap of a few GHz~\cite{ultrafastiontwo,RWA}, so there are rooms for improvement of the clock speed. It is therefore an important question whether the hyperfine state transitions can be driven with an extremely fast optical means, such as femtosecond laser pulses, which is the subject of this paper.

The time scale of ultrafast optical interactions is on the order of less than a picosecond, much shorter than the phase evolution time of atomic hyperfine states with an energy gap of a few GHz. So, from the interaction Hamiltonian,
\begin{equation}
H_{\rm int}=-\hat{\bf \mu}\cdot {\bf E}+A_{\rm FS}{\bf L}\cdot {\bf S}+\cancel{A_{\rm HFS}{\bf J}\cdot {\bf I}},
\end{equation}
where the interactions are electric dipole, fine-structure, and hyperfine interactions, respectively, 
the hyperfine interaction (the last term) can be ignored and thus, the nuclear degree of freedom is unchanged. In other words, subpicosecond-time-scale optical interactions are local unitary operations acting only on the electronic subspace of the atom, {i.e.}, $\hat{U}_{\rm ultrafast}=\hat{U}_{\bf J} \otimes \hat{\mathbbm{1}}_{\bf I} $. On the other hand, atomic hyperfine clock states are maximally entangled states of the electronic and nuclear degrees of freedom, and their entire Hilbert space should be accessible with local operations and classical communications~\cite{entreview}. Therefore, it is possible to achieve hyperfine-state qubit gates using only subpicoseond optical local unitary operations, as long as they preserve entanglement (without resorting to classical communications).

In this paper, we propose and provide a proof-of-principle experimental demonstration of subpicosecond-time-scale $X$ rotations of atomic hyperfine clock-state qubits. We first describe the subpicosecond optical transition of atoms between fine-structure states and its effect on the hyperfine clock states, and then how to use this effect to implement an $X$ rotation of the clock-state qubit in Sec.~II. After a brief experimental procedure in Sec.~III, the theoretical analysis for the subpicosecond optical transitions is presented in Sec.~IV (details are provided in Appendix). Experimental results are presented in Sec.~V, followed by discussions on maximally possible gate fidelity, universal single qubit gate scheme, and estimated overall computational speedup in Sec. VI, before the conclusion in Sec.~VII.

\section{Clock state $X$ rotations} \label{theory}
\begin{figure}[t]
\centering
\includegraphics[width=0.43\textwidth]{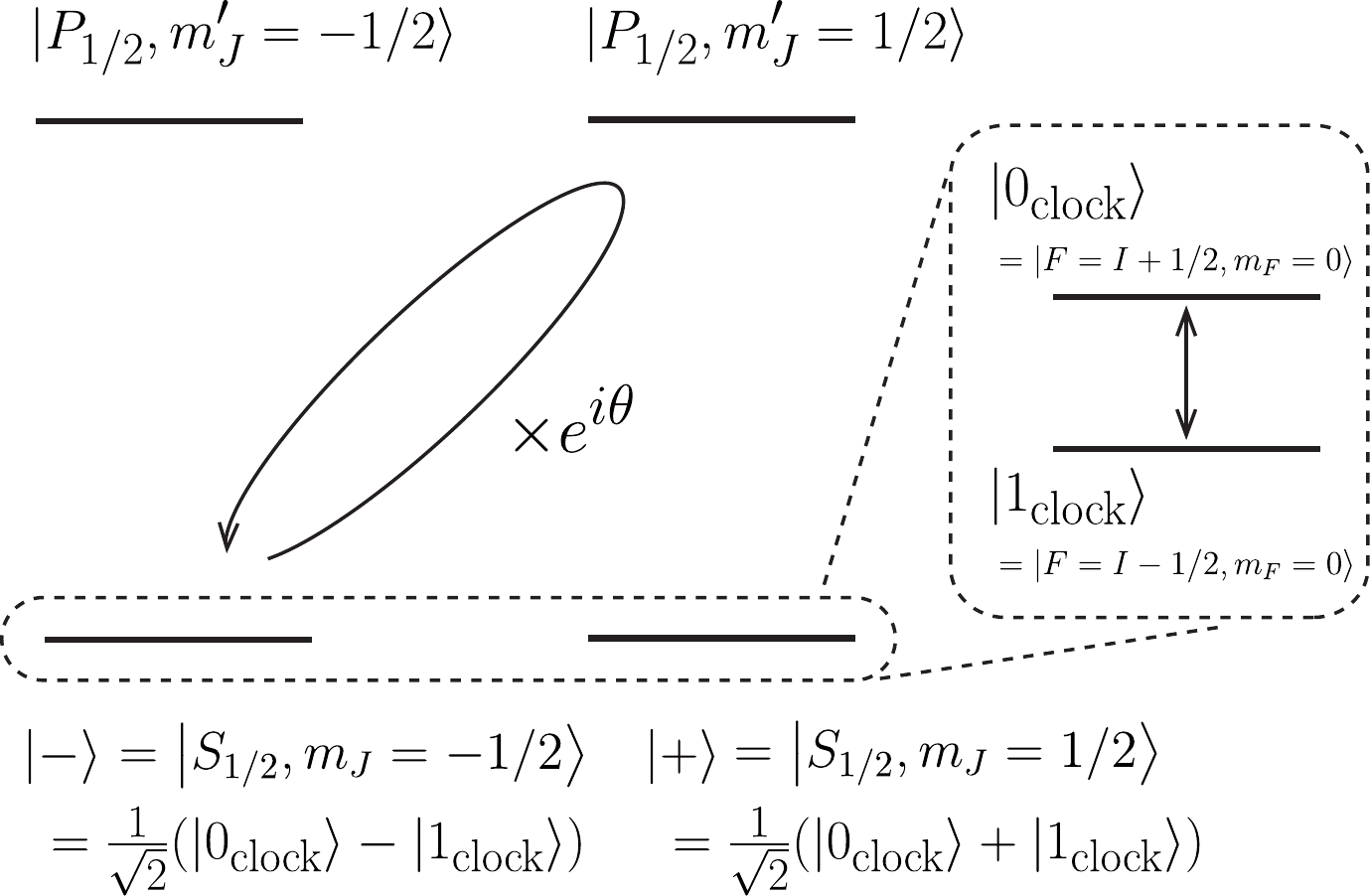}
\caption{Subpicosecond $X$ rotation scheme between the clock states, $\ket{0_{\rm clock}}=\ket{S_{1/2}, F=I+1/2, m_F=0}$ and $\ket{1_{\rm clock}}=\ket{S_{1/2}, F=I-1/2, m_F=0}$, where a circularly polarized ($\sigma^+$) laser pulse drives the cyclic Rabi oscillation of the ground fine-structure sublevel $\ket{-}$ through the fine-structure excited state $\ket{P_{1/2}, m_J'=1/2}$. While, the other ground fine-structure sublevel $\ket{+}$ remains as a dark state due to the selection rule, resulting in the relative phase $\theta$ between $\ket{\pm}$. Note that a $\pi$-polarized pulse cannot make this kind of qubit rotations: because the linear polarized light (the equal superposition of $\sigma^\pm$-polarized lights) drives the same Rabi oscillations to both $\ket\pm$ (so, no relative phase). }
\label{fig1}
\end{figure}
The atomic clock states are the ground hyperfine states of $m_F=0$ and different two $F=I\pm1/2$; for example,  $\ket{0_{\rm clock}}\equiv \ket{5S_{1/2}, F=2, m_F=0}$ and $\ket{1_{\rm clock}} \equiv \ket{5S_{1/2}, F=1, m_F=0}$ in $^{87}$Rb. These clock states have a radio frequency energy separation, so in a subpicosecond time scale (where the nuclear degree of freedom is frozen) an optical transition is not conveniently described with the hyperfine state basis. Therefore, we first use the fine-structure basis, which consist of the ground and excited states with an optical frequency energy gap, and then utilize the angular-momentum relations~\cite{angmom} between the ground fine-structure sublevels $\ket{S_{1/2}, m_J=\pm 1/2}$ and the clock states, i.e., $ \ket{S_{1/2}, m_J=\pm 1/2}=({\ket{0_{\rm clock}} \pm \ket{1_{\rm clock}}})/\sqrt{2}\equiv \ket{\pm}$. (The nuclear degree of freedom $I$ is omitted since it is frozen.) 

These fine-structure sublevels, $\ket{\pm}$, are the eigenstates of the Pauli $X$ operator in the clock-state Hilbert space of $\ket{\psi}=\alpha \ket{0_{\rm clock}}+ \beta \ket{1_{\rm clock}}$. Therefore, if there is an optical means that directly induces the relative phase between $\ket{\pm}$, we achieve general $X$ rotations of $\ket{\psi}=(\alpha-\beta) \ket{-}+ (\alpha+\beta) \ket{+}$. Figure~\ref{fig1} shows a $2\pi$ Rabi oscillation (the optical excitation and return) induced by a $\sigma^+$-polarized laser pulse, where the $\ket{-}$ gains the geometric phase $\theta$~\cite{geophase} while the $\ket{+}$ is intact, i.e., $\ket{\psi} \rightarrow (\alpha-\beta) e^{i\theta} \ket{-}+ (\alpha+\beta) \ket{+}$ (the $X$ rotation). The final state is given by $\hat{R}_{\hat{x}}(\theta)\ket{\psi}$ with the rotation matrix 
\begin{equation}
\hat{R}_{\hat{x}}(\theta) =
\begin{pmatrix}
\cos \frac{\theta}{2} & - i \sin \frac{\theta}{2} \\
-i \sin \frac{\theta}{2} &  \cos \frac{\theta}{2}
\end{pmatrix}e^{i\theta/2},
\end{equation}
in the clock-state basis $\{\ket{0_{\rm clock}}, \ket{1_{\rm clock}} \}$, where the global phase $e^{i\theta/2}$ is ignored because it plays no role.  

The angle $\theta$ is given as a function of detuning $\Delta=\omega-\omega_0$, where $\omega$ and $\omega_0$ are the laser frequency and atomic resonance, respectively. For example, in the Rosen-Zener model case~\cite{geophasegate}, the angle $\theta$ is obtained as $\theta=2\arctan({\Omega_{RZ}}/{\Delta})$ where $\Omega_{RZ}$ is the peak Rabi frequency. In our experimental condition to be described in the   next section, an analytical expression for $\theta$ is not available (due to the dynamic Stark shift from D$_2$ transition and the pulse shape), but we will show through numerical calculation that the full range of $\theta$ can be achieved as a function of $\Delta$.
\\ \\

\section{Experimental Procedure} \label{results}
Experimental investigation of the subpicosecond $X$ rotation of the atomic clock states was performed in a setup described in our early works~\cite{Lim,  LeeOL2015, HSKPRA2015}. The setup composed of  a magneto-optical trap (MOT)~\cite{MOT} for cold rubidium atoms ($^{87}$Rb) and a femtosecond laser amplifier~\cite{amp}. The laser was operated with 1~kHz repetition rate and we used a programmable acousto-optic modulator (ADPDF, Dazzeler from Fastlite)~\cite{AOPDF} to produce temporal Gaussian pulses with a center frequency detuned up to 1.5~THz from the D$_1$ resonance (795~nm). Each shaped pulse had a pulse-bandwidth of 2.5~THz (FWHM) and a pulse-energy up to 5~$\mu$J, enough to perform up to $4\pi$ Rabi oscillation, when being focused with a lens of 500~mm focal length. 

After being cooled in the MOT and depumped from $F$=2 to $F$=1, the atoms were prepared in an equal mixture of the clock state $\ket{1_{\rm clock}}$ and the other two magnetic sublevels of the ground $F=1$ hyperfine state, i.e., $\hat\rho_{\rm init}=\sum_{m_F} \frac{1}{3}\ket{5S_{1/2},F=1,m_F}\bra{5S_{1/2}, F=1,m_F}$. Then, the laser pulse interacted with the atoms to drive the D$_1$ transition $\ket{5S_{1/2}} \leftrightarrow \ket{5P_{1/2}}$. After 30~$\mu$s (for complete spontaneous emission of remaining $\ket{5P_{1/2}}$ population), the $F$=2 state population was measured through absorption imaging (10$\mu$s exposure, $F=2 \rightarrow F'=3$ D$_2$ transition) as a function of the pulse-area~\cite{McCall-Hahn}, $\mathcal{A}=\int_{-\infty}^{\infty} \Omega(t) dt$ with Rabi frequency $\Omega$ and detuning $\Delta$.

\section{Transition probabilities} \label{results}

We first consider the transition probability between the ground hyperfine sublevels, from $\ket{5S_{1/2}, {F=1}, m_F}=\ket{1,m_F}$ to $\ket{5S_{1/2}, {F=2}, m_F''}=\ket{2,m_F''}$ via $\ket{5P_{1/2},m_J'}$ (D$_1$ transition). The given probability is the sum of the direct optical transition (qubit rotation) from $\ket{1, m_F}$ to $\ket{2, m_F''=m_F}$ and the spontaneous emission from the excited fine-structure states $\ket{5P_{1/2}, m_J'}$.

The case of $\sigma^+$ polarization is given as 
\begin{widetext}
\begin{eqnarray}
P(\mathcal{A},\Delta;\sigma^+,m_F)&=&
\abs{\bra{2,m_F} \hat U \ket{1,m_F}}^2 
+ P_{\rm se}(1/2;\sigma^+,m_F)\abs{\bra{5P_{1/2},1/2}\hat U\ket{1,m_F}}^2 \nonumber\\
&=&A_{m_F} \abs{1-{\bra{5S_{1/2},-1/2} \hat{U}\ket{5S_{1/2},-1/2}} }^2+ B_{m_F}\abs{ \bra{5P_{1/2}, 1/2} \hat{U} \ket{5S_{1/2}, -1/2}}^2,
\label{a_eq1}
\end{eqnarray}
where $\hat U(\mathcal{A},\Delta;\sigma^+)$ is the subpicosecond optical interaction, given as a function of pulse-area $\mathcal{A}$ and detuning $\Delta$, and  $P_{\rm se}(m_J';\sigma^+,m_F)$ is the conditional probability of spontaneous emission from $\ket{5P_{1/2},m_J'}$ to all magnet sublevels $\ket{F=2, m_F''}$ ($m_F''={\pm2},\pm1,0$), if the initial state is $\ket{1,m_F}$.  Using the Clebsch-Gordan coefficients~\cite{angmom} between the $\ket{J, m_J}$ and $\ket{F, m_F}$ bases, we get for $m_F=-1,0,1$, respectively,
\begin{eqnarray}
A_{m_F} &=& \abs{C^{\frac{1}{2},\frac{3}{2},2}_{\frac{1}{2},m_F- \frac{1}{2}} C^{\frac{1}{2},\frac{3}{2},1}_{\frac{1}{2},m_F- \frac{1}{2}} }^2 = \frac{3}{16},\frac{1}{4},\frac{3}{16} \\
B_{m_F} &=& \sum_{F'=1,2,q=0,\pm 1}  \abs{C^{\frac{1}{2},\frac{3}{2},F'}_{\frac{1}{2},m_F+\frac{1}{2}} D^{F',1,2}_{m_F+1,q}}^2 \abs{ C^{\frac{1}{2},\frac{3}{2},1}_{-\frac{1}{2},m_F+\frac{1}{2}} }^2 =  \frac{1}{6}, \frac{7}{24}, \frac{3}{8},
\end{eqnarray}
where $D^{j_1,j_2,j_3}_{m_1,m_2}$ = $\sqrt{2(2j_1+1)}\left\{\begin{matrix} 1/2 & 1/2 & 1 \\ j_1 & j_3 & 3/2\end{matrix}\right\}C^{j_1,j_2,j_3}_{m_1,m_2}$, with curly brackets denoting the Wigner 6-j symbol (See Appendix for details). On account of symmetry, the $\sigma^+$ and $\sigma^-$ polarization cases are the same, except that $m_J$ and $m_F$ are replaced by $-m_J$ and $-m_F$, respectively.

The $\pi$ polarization case is given by
\begin{equation}
P(\mathcal{A},\Delta;\pi,m_F) = \sum_{m_J'=\pm1/2} P_{\rm se}(m_J';\pi,m_F) \abs{\bra{5P_{1/2},m_J'}\hat U\ket{1,m_F}}^2   = {E_{m_F}} \abs{ \bra{5P_{1/2}, -1/2} \hat{U} \ket{5S_{1/2}, -1/2}}^2,
\label{a_eq2}
\end{equation}
where the coefficient is given for $m_F=-1,0,1$, respectively, by 
\begin{equation}
E_{m_F}=\sum_{\substack{F'=1,2, q=0,\pm 1,\\ m_J=\pm\frac{1}{2}}}  \abs{C^{\frac{1}{2},\frac{3}{2},F'}_{m_J,m_F-m_J} D^{F',1,2}_{m_F,q}}^2 \abs{ C^{\frac{1}{2},\frac{3}{2},1}_{m_J,m_F-m_J} }^2 
=\frac{7}{12}, \frac{1}{2}, \frac{7}{12}.
\end{equation}
\end{widetext}
Note that there is only spontaneous emission contribution, that is proportional to the transition between the ground and excited fine-structure states, with no direct $\ket{1,m_F}\rightarrow \ket{2,m_F}$ hyperfine transition, i.e., $\abs{\bra{2,m_F} \hat U \ket{1,m_F}}^2=0$ (See Appendix for details). 

\begin{figure}[H]
\centering
\includegraphics[width=0.45\textwidth]{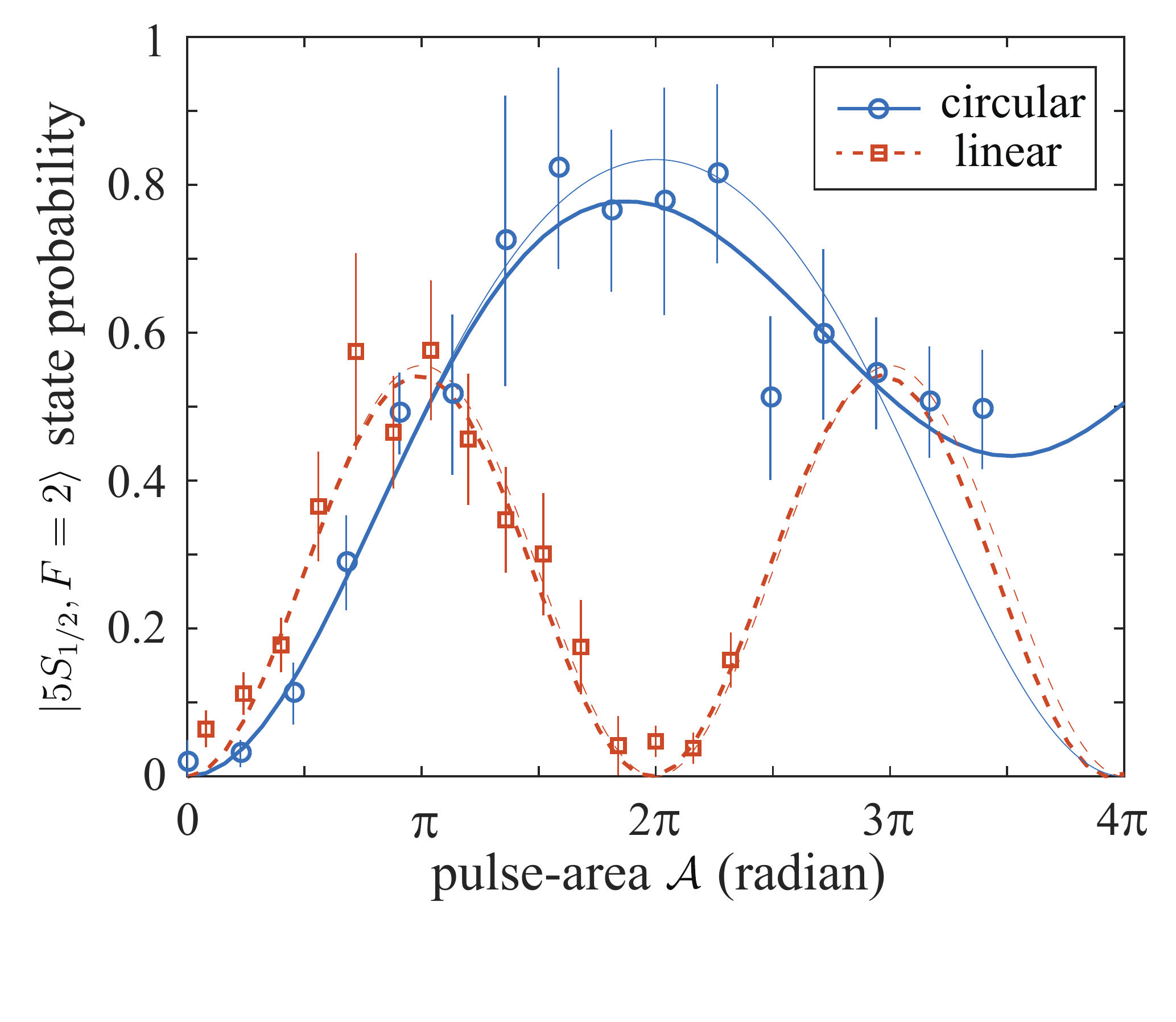}
\caption{(Color online) $|5S_{1/2}, {F=2}\rangle$ state probability $P(\mathcal{A})$ vs. pulse-area $\mathcal{A}$ for $\sigma^+$ and $\pi$ resonant D$_1$ transitions, respectively, where blue circles ($\sigma^+$) and red squares ($\pi$) represent the experimental data. The thick solid and dashed lines are the corresponding numerical calculations, with the thin lines showing the calculated results for the weak excitation regime. The error bars indicate the standard error of mean.}
\label{fig2}
\end{figure}

\begin{figure*}[tbh]
\centering
\includegraphics[width=0.9\textwidth]{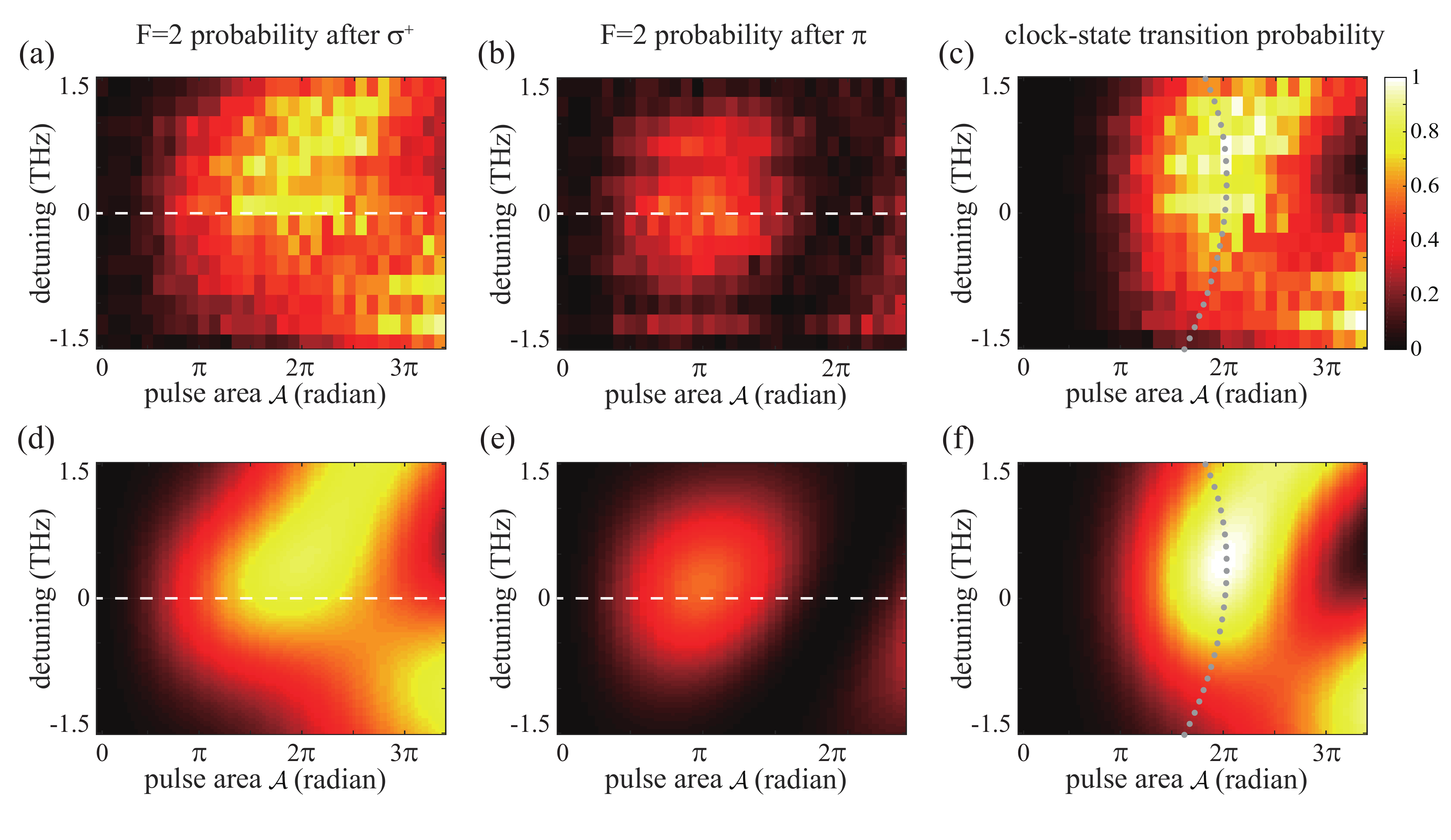}
\caption{(Color online) (a-b) Experimental results for the population of $5S_{1/2}, F=2$ states after $\sigma^+$ (a) and $\pi$ (b) transitions as a function of pulse-area $\mathcal{A}$ and detuning $\Delta/2\pi$. (c) Transition probability of the clock-state qubit ($m_F=0$ only), retrieved using (a), (b), and Eq.~(3). (d-f) TDSE results corresponding to (a-c).}
\label{fig3}
\end{figure*}

\begin{figure*}[thb]
\centering
\includegraphics[width=0.8\textwidth]{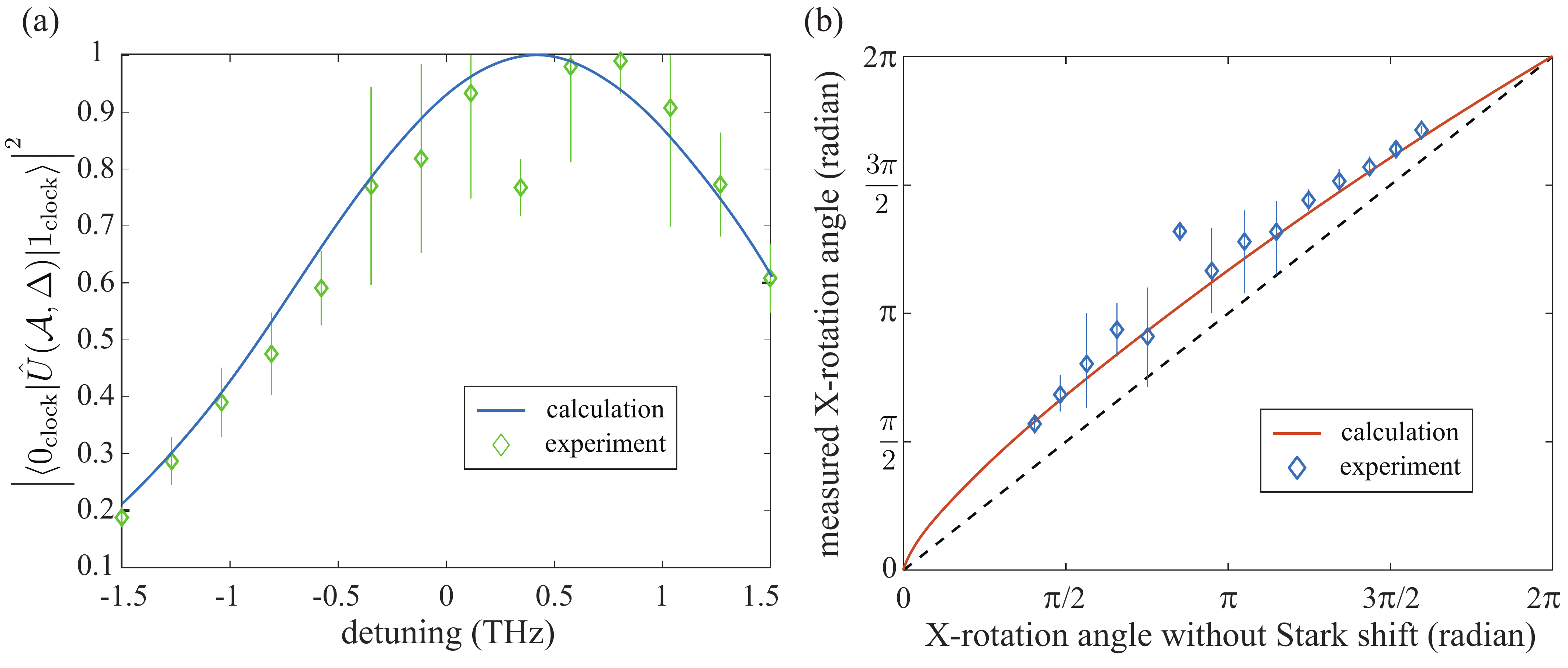}
\caption{(Color online) (a) Qubit transition probability, experimentally retrieved for only clock state contribution (green diamonds), and the corresponding calculation results (blue solid line) plotted as a function of detuning $\Delta/2\pi$. (b) Measured $X$-rotation angles (blue diamonds) and full calculation (red solid line) compared with the two-level model calculation ($x$-axis). The errorbars indicate the standard error of mean. The experimental data are from the complete population return points (dotted line) in Figs.~\ref{fig3}(c) and \ref{fig3}(f).}
\label{fig4}
\end{figure*}

\section{Results} \label{results}

In our experiment, the initial state is the mixed state of $F$=1 states, or $ \hat\rho_{\rm init}= \frac{1}{3}\sum_{m_F} \ket{1,m_F}\bra{1,m_F}$,
so the final $F$=2 state probability is given by
\begin{eqnarray} 
P(\mathcal{A},\Delta;\sigma^+) 
&=& \frac{5}{24} \abs{1-{\bra{5S_{1/2},-1/2} \hat{U}\ket{5S_{1/2},-1/2}} }^2 \nonumber \\ 
&+& \frac{5}{18}
\abs{ \bra{5P_{1/2}, 1/2} \hat{U} \ket{5S_{1/2}, -1/2}}^2, \label{eq8} \\
P(\mathcal{A},\Delta;\pi) 
&=& \frac{5}{9} \abs{ \bra{5P_{1/2}, -1/2} \hat{U} \ket{5S_{1/2}, -1/2}}^2
\label{eq9}
\end{eqnarray}
which can be obtained from Eqs.~\eqref{a_eq1} and \eqref{a_eq2} by replacing $A_{m_F}$, $B_{m_F}$, and $E_{m_F}$ with $\frac{1}{3}\sum_{m_F}A_{m_F}$, $\frac{1}{3}\sum_{m_F}B_{m_F}$, and $\frac{1}{3}\sum_{m_F}E_{m_F}$, respectively, since each ground and excited states pair for each $m_F$ forms independent two-state system and their dynamic behaviors in fine-structure basis are all the same.

Figure~\ref{fig2} shows the $F$=2 state probabilities, $P(\mathcal{A};{\sigma^+})$ and $P(\mathcal{A};{\pi})$, measured for $\sigma^+$ and $\pi$ transitions, respectively, under the resonant excitation condition $\Delta=0$. In this condition, $\hat U$ describes the resonant Rabi oscillation between the ground and excited fine-structure states. $P(\mathcal{A};{\sigma^+})$ is given as the sum of the hyperfine-state rotation ($\ket{F=1} \leftrightarrow \ket{F=2}$) and the spontaneous emission from the population remaining in $\ket{5P_{1/2}}$. $P(\mathcal{A};{\pi})$ has only the latter contribution, the fine-structure Rabi oscillation profile, because $\pi$-polarized light induces no hyperfine-state rotation. In the weak-excitation regime (where the dynamic Stark shift is negligible), their analytic forms are obtained as 
\begin{align}
&P(\mathcal{A};\sigma^+) = \frac{5}{24} \left(1-\cos\frac{\mathcal{A}}{2}\right)^2 +  \frac{5}{18} \sin^2\frac{\mathcal{A}}{2}, \label{resonance}\\
&P(\mathcal{A};\pi) = \frac{5}{9} \sin^2\frac{\mathcal{A}}{2},
\end{align}
as shown in Fig.~\ref{fig2}. When the dynamic Stark-shift (involving 5P$_{3/2}$, 5D$_{3/2}$, and 5D$_{5/2}$) is taken into account, the numerical calculation of Eqs.~(\ref{eq8}, \ref{eq9}) using the time-dependent Sch\"odinger equation (TDSE) results in a good agreement with the experiment. Note that $\max\left(P(\mathcal{A};\sigma^+)\right)$ is $5/6\approx83\%$ at the complete population return ($\mathcal{A}=2\pi$) in Eq.~\eqref{resonance}, which corresponds to 1 (perfect population transfer) if the initial state is the clock state $\ket{1_{\rm clock}}$ with $A_{m_F=0}$ rather than $\frac{1}{3}\sum_{m_F}A_{m_F}$.

We now probe these probabilities by changing both $\mathcal{A}$ and $\Delta$, i.e., $P(\mathcal{A},\Delta;{\sigma^+})$ and $P(\mathcal{A},\Delta;{\pi})$, with results shown in Figs.~\ref{fig3}(a) and (b). 
The dashed lines at the resonance condition $\Delta=0$ in the figure correspond to Fig.~\ref{fig2}. The clock-state rotation (only) contribution Fig.~\ref{fig3}(c) is extracted from the data in Fig.~\ref{fig3}(a), while the pulse-area ($x$-axis) is calibrated with the data (along the $\Delta=0$ line) from Figs.~\ref{fig3}(b). Here, we use the fact that the initial state is an equal mixture of the magnetic sublevels and that each ratio among them for the hyperfine-state rotation and also the spontaneous emission are known. After the spontaneous emission is subtracted from Fig.~\ref{fig3}(a), $A_{m_F=0}/(\frac{1}{3}\sum_{m_F}A_{m_F})$ factor is multiplied to obtain the clock state contribution. The corresponding numerical calculations depicted in Figs.~\ref{fig3}(d), (e), and (f) show a good agreement with the experimental results.

Using these measurements, we can retrieve the qubit $X$-rotation performance. The result is shown in Fig.~\ref{fig4}. The qubit transition probability from $\ket{1_{\rm clock}}$ to $\ket{0_{\rm clock}}$ is shown in Fig.~\ref{fig4}(a), where the data is from the dotted line in Fig.~\ref{fig3}(c) that corresponds to the complete population return to the ground states, i.e., {$\hat{U}\ket{1_{\rm clock}}= -i \sin {\theta}/{2} \ket{0_{\rm clock}}+ \cos{\theta}/{2} \ket{1_{\rm clock}}$.} Also, in Fig.~\ref{fig4} (b), the corresponding $X$-rotation angles $\theta$ are extracted and plotted as a function of weak-excitation regime calculations ($x$-axis). Compared to the TDSE calculation result (solid line), the experimental data (diamonds) shows good agreement. The discrepancy between the TDSE and weak-excitation regime results (dashed line) is mainly due to the dynamic Stark shift. Our experiment performed at 350~fs results in over 98\% transition probability, $\abs{\bra{0_{\rm clock}} \hat{U}(\mathcal{A}, \Delta=2\pi\times0.52~{\rm THz}) \ket{1_{\rm clock}}}^2>98\%$ in Fig.~4(a).

\section{Discussions} \label{results}

\begin{figure}[tbh]
\centering
\includegraphics[width=0.45\textwidth]{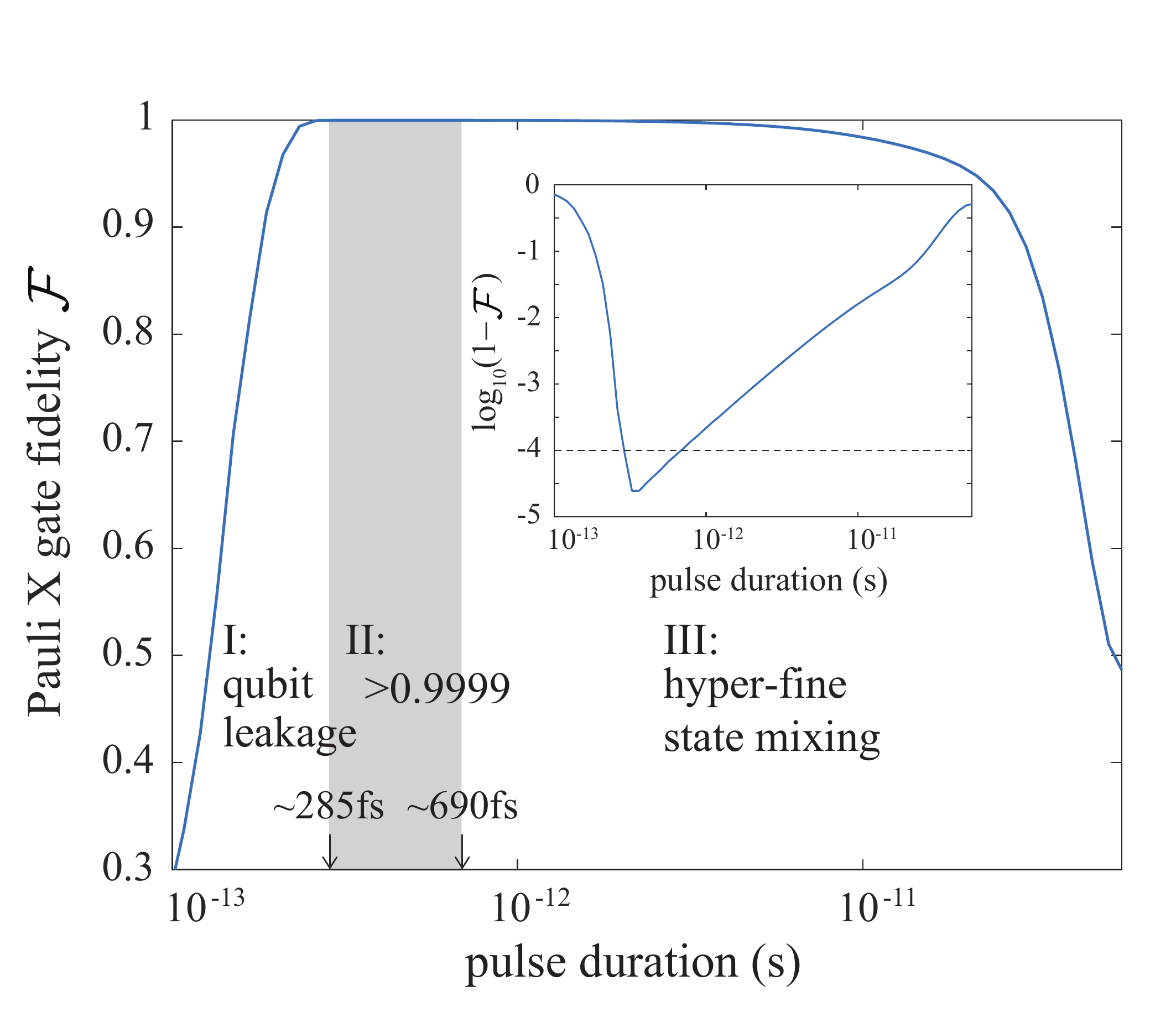}
\caption{(Color online) Pauli $X$ gate fidelity $\mathcal{F}$ vs. laser-pulse duration: The population transfer, $\abs{\bra{0_{\rm clock}} \hat{U}(\mathcal{A}, \Delta) \ket{1_{\rm clock}}}^2$, between hyperfine clock states is optimized with pulse-area $\mathcal{A}$ and detuning $\Delta$, as varying the pulse duration. At the optimal points $\mathcal{A}^*$ and $\Delta^*$ for each pulse duration ($x$-axis, log scale), the numerical calculation results of the $\hat{X}$ gate fidelity $\mathcal{F}=\overline{\abs{\bra{\psi_{\rm in}} \hat{X}^\dagger\hat{U}(\mathcal{A}^*, \Delta^*) \ket{\psi_{\rm in}}}^2}$ are shown, and the region with over 0.9999 fidelity is shaded.
}
\label{fig5}
\end{figure}

We can also investigate the maximally possible Pauli $X$ gate fidelity of our scheme through numerical simulation. Figure~\ref{fig5} shows the TDSE calculation for $\mathcal{F}=\overline{\abs{\bra{\psi_{\rm in}} \hat{X}^\dagger\hat{U}(\mathcal{A}^*, \Delta^*) \ket{\psi_{\rm in}}}^2}$, where the gate fidelity $\mathcal{F}$ is defined as an average over the set of all input states~\cite{gatefidelity}, i.e., $\ket{\psi_{\rm in}} \in \{\ket{0_{\rm clock}},\ket{1_{\rm clock}},\ket+,(\ket{0_{\rm clock}}+i\ket{1_{\rm clock}})/\sqrt{2}\}$. The given result is obtained through varying the pulse duration ($x$-axis) while the optimal pulse-area and detuning, $\mathcal{A}^*$ and $\Delta^*$, are chosen to maximize $\mathcal{F}$ at each pulse duration.  The region of high fidelity is identified between about 285 and 690~fs (region II). When the pulse duration is too short (region I), transitions to other states (mainly $5P_{3/2}$, $5D_{3/2}$, and $5D_{5/2}$) are not negligible. When the pulse duration is too long (region III), nuclear motion is no longer frozen during the interaction and, thus, additional $Z$ rotations by hyperfine evolution cause state mixing of the qubit. Within the subpicosecond time scale (region II, shaded in Fig. \ref{fig5}), our scheme theoretically predicts a high fidelity of over 99.99\%.


The universal single qubit control requires an extra noncommuting rotation gate besides $X$ rotations~\cite{NC}. If the time scale is limited in the subpicosecond range, where the interaction about the nuclear spin degree of freedom is frozen, the possible gates are only $X$ gates and not generalized to $Y$ or $Z$ gates. This can be understood based on entanglement arguments about local unitary operations~\cite{entreview}. An arbitrary qubit rotation $\hat{R}_{\hat{n}}(\theta)\ket{\psi}$ is given by
\begin{equation}
\hat{R}_{\hat{n}}(\theta)=
\begin{pmatrix}
\cos\frac{\theta}{2}-in_z\sin\frac{\theta}{2} & -(n_y+in_x)\sin\frac{\theta}{2} \\
(n_y-in_x)\sin\frac{\theta}{2} & \cos\frac{\theta}{2}+in_z\sin\frac{\theta}{2}
\end{pmatrix},
\end{equation}
where $\hat{n}=(n_x, n_y, n_z)$ and $\theta$ are the rotational axis and angle, respectively. Considering the fact that the bipartite entanglement between electronic and nuclear degrees of freedom is preserved during a subpicosecond optical cyclic evolution, we calculate the entanglement entropy~\cite{ententropy}  of the rotated state,
\begin{equation}
E\left(R_{\hat{n}}(\theta)\ket{\psi_{\rm clock}}\right)=- P_+\log_2 P_+ - P_-\log_2 P_-,
\end{equation}
with $P_+ = \frac{1}{2} |(\alpha+\beta)(\cos\frac{\theta}{2}-in_x\sin\frac{\theta}{2})+(\alpha-\beta)(n_y-in_z)\sin\frac{\theta}{2}|^2$ 
and $P_-=\frac{1}{2} |(\alpha-\beta)(\cos\frac{\theta}{2}+in_x\sin\frac{\theta}{2})-(\alpha+\beta)(n_y+in_z)\sin\frac{\theta}{2}|^2/{2}$. The condition, necessary for preserved entanglement, is obtained as $n_x=1$ and $n_y=n_z=0$ for all $\theta$. Thus, the only gate available for subpicosecond optical interactions is the $X$ rotation, $R_{\hat{x}}(\theta)$. In order to achieve universal single qubit controls, it is necessary to use the hyperfine interaction in a longer time scale. The hyperfine interaction is properly described in the clock-state qubit basis $\{\ket{0'_{\rm clock}},\ket{1'_{\rm clock}}\}$ in the interaction picture, where $\ket{0'_{\rm clock}}=\ket{0_{\rm clock}}$, $\ket{1'_{\rm clock}}=e^{-i\omega_{\rm hf}t}\ket{1_{\rm clock}}$, and $\omega_{\rm hf}$ is the ground hyperfine splitting (about $2\pi\times6.8$~GHz for $^{87}$Rb). Then, the ground fine-structure states $\ket{\pm}=\ket{0'_{\rm clock}}\pm e^{i\omega_{\rm hf} t}\ket{1'_{\rm clock}}$ are no longer eigenstates of the Pauli $X$ operator in general, and become the eigenstates of the Pauli $Y$ operator after the quarter hyperfine period, $\pi/2\omega_{\rm hf}$. Therefore, the time-delayed $X$ rotation becomes a $Y$ rotation, allowing two distinct single-qubit rotations sufficient for general single-qubit rotations. This method could be implemented with pulse shaping technique~\cite{pulseshaping} or optical frequency combs~\cite{freqcomb}.

An overall speedup of a particular quantum computation scheme can be estimated using Amdahl's law~\cite{Amdahl}. The speedup $S$ from improving a part (single qubit gates in our scheme) of the whole system (universal gate set = single qubit gates + CNOT) is theoretically given by 
\begin{equation}
S=\frac{1}{(1-r_p)+{r_p}/{n}},
\end{equation}
where $r_p$ is the ratio of the improved part in the whole system and $n$ is the speedup of the part. Our scheme suggests an improved operation speed for all single qubit gates, by {a large factor ($>10^3$) compared to the conventional approaches~\cite{microwave,RB,3D,MI,ionMW,Raman,Raman2,micromirror,ionRaman,ionRaman2}}, while there is no improvement for CNOT. All quantum algorithms can be decomposed by two-qubit controlled unitary operations $\Lambda(\hat U)=\ket0\bra0\otimes\hat{\mathbbm{1}}+\ket1\bra1\otimes\hat U$ which consist of three single-qubit gates and two CNOTs~\cite{Preskill}. Under the assumption of equal appearance of these five gates, we get the ratio of single-qubit gate duration $r_p=15/16$ and $3/23$,  respectively, microwave~\cite{microwave,RB,3D,MI} and Raman~\cite{Raman,Raman2,micromirror} schemes in Rydberg-based neutral atom platforms~\cite{QCreview, Ry,Ry2,Ry3,Ry4}.  Also, $r_p=3/23$ for ion trap platforms, where two-qubit gate is typically over an order slower than single-qubit gate~\cite{iontrap,ionMW,ionRaman,ionRaman2}. Then, it is expected that our scheme achieves about 1500\% (microwave) and 15\% (Raman) of overall speedup improvement in neutral atoms, and 15\% in ions, respectively.

\section{Conclusion} \label{conclusion}
In summary, we have demonstrated a population transfer between atomic hyperfine states using a single laser-pulse in the femtosecond time scale. The reason how in our scheme an optical pulse with a few THz bandwidth can control the qubit system with an energy splitting of a few GHz---which seemingly violates the quantum speed-limit theorem~\cite{speedlimit}---is the selection rule imposed to subpicosecond optical interactions that negligibly change the nuclear degree of freedom. Utilizing the geometric and dynamic phases induced to the hyperfine states during the subpicosecond optical Rabi oscillation, we can conclude that $X$ rotations of atomic clock states can be implemented. Alternatively, more direct demonstrations such as pure state preparation, $m_F$-selective measurements in a trapped single-atom system may be possible.
\\
\begin{acknowledgements}
This research was supported by Samsung Science and Technology Foundation [SSTF-BA1301-12]. We thank the anonymous referee for suggesting Amdahl's law to estimate the overall computation speedup.
\end{acknowledgements}

\appendix
\section*{Appendix: Derivation of $A_{m_F}$, $B_{m_F}$, $E_{m_F}$}

The subpicosecond optical transition of the rubidium atoms from an initial $F=1$ ground hyperfine sublevel to the $F=2$ sublevels is obtained. The initial and final states are written as a superposition of the fine-structure magnetic sublevels,  given respectively by
\begin{widetext}
\begin{eqnarray}\label{rel1}
\ket{5S_{1/2}, F=1,m_F}
&=&\sum_{\substack{m_J=\pm1/2}}C^{\frac{1}{2},\frac{3}{2},1}_{m_J,m_F-m_J}\ket{5S_{1/2},m_J}\ket{I=3/2,m_F-m_J} \equiv \ket{1, m_F},
\end{eqnarray}
\begin{eqnarray}\label{rel2}
\ket{5S_{1/2}, F=2,m_F''}
&=&\sum_{\substack{m_J''=\pm1/2}}C^{\frac{1}{2},\frac{3}{2},2}_{m_J'',m_F''-m_J''}\ket{5S_{1/2},m_J''}\ket{I=3/2,m_F''-m_J''}\equiv \ket{2, m_F''}.
\end{eqnarray}

{\it The $\sigma$ polarization case :}
The transition probability $P(\mathcal{A},\Delta;\sigma^+,m_F)$ is given as the first line in Eq.~\eqref{a_eq1}, which is the sum of the direct optical transition from $\ket{1,m_F}$ to $\ket{2, m_F''}$ and the spontaneous emission from $\ket{5P_{1/2}, m_J'}$ to $\ket{2, m_F''}$, i.e. 
\begin{align}\label{tprob1}
&P(\mathcal{A},\Delta;\sigma^+,m_F)\nonumber\\
&=\sum_{m_F''={\pm2},\pm1,0}\abs{\bra{2,m_F''} \hat U \ket{1,m_F}}^2 + \sum_{\substack{m_J'=\pm1/2,\\m_I'={\pm3/2},\pm1/2}}P_{\rm se}(m_J';m_F,\sigma^+)\abs{\bra{5P_{1/2},m_J'}\bra{I=3/2,m_I'}\hat U\ket{1,m_F}}^2\nonumber\\
&=\abs{\bra{2,m_F} \hat U \ket{1,m_F}}^2 + P_{\rm se}(1/2;m_F,\sigma^+)\abs{\bra{5P_{1/2},1/2}\bra{I=3/2,m_F+1/2}\hat U\ket{1,m_F}}^2.
\end{align}
The first term (the direct optical transition) is obtained as $\abs{\bra{2,m_F} \hat U \ket{1,m_F}}^2$, because
\begin{align} \label{eq18}
&\bra{2,m_F''} \hat U \ket{1,m_F}\nonumber\\
&=\sum_{\substack{\\m_J''=\pm1/2,\\m_J=\pm1/2}}C^{\frac{1}{2},\frac{3}{2},2}_{m_J,m_F''-m_J''}C^{\frac{1}{2},\frac{3}{2},1}_{m_J,m_F-m_J}\bra{5S_{1/2},m_J''}\hat U \ket{5S_{1/2},m_J}\langle{I=3/2,m_F''-m_J''}|{I=3/2,m_F-m_J}\rangle\nonumber\\
&=\sum_{\substack{m_J''=\pm1/2,\\m_J=\pm1/2}}C^{\frac{1}{2},\frac{3}{2},2}_{m_J,m_F''-m_J''}C^{\frac{1}{2},\frac{3}{2},1}_{m_J,m_F-m_J}\bra{5S_{1/2},m_J}\hat U \ket{5S_{1/2},m_J}\delta_{m_J'',m_J}\langle{I=3/2,m_F''-m_J}|{I=3/2,m_F-m_J}\rangle\nonumber\\
&=\bra{2,m_F} \hat U \ket{1,m_F}\delta_{m_F'',m_F},
\end{align}
where we use the facts $\hat{U}=\hat{U}_{\bf J} \otimes \hat{\mathbbm{1}}_{\bf I}$ (independent of $I$) and the dipole selection rule $\bra{5S_{1/2},m_J''}\hat U \ket{5S_{1/2},m_J}=\bra{5S_{1/2},m_J}\hat U \ket{5S_{1/2},m_J}\delta_{m_J'',m_J}$. Further, Eq.~\eqref{eq18} can be simplified as
\begin{equation}
\bra{2,m_F} \hat U(\mathcal{A},\Delta;\sigma^+) \ket{1,m_F}=C^{\frac{1}{2},\frac{3}{2},2}_{\frac{1}{2},m_F-\frac{1}{2}}C^{\frac{1}{2},\frac{3}{2},1}_{\frac{1}{2},m_F-\frac{1}{2}} +C^{\frac{1}{2},\frac{3}{2},2}_{-\frac{1}{2},m_F+\frac{1}{2}}C^{\frac{1}{2},\frac{3}{2},1}_{-\frac{1}{2},m_F+\frac{1}{2}}\bra{5S_{1/2},-1/2} \hat{U}\ket{5S_{1/2},-1/2},
\end{equation}
because of the selection rule  $\bra{5S_{1/2},1/2}\hat U \ket{5S_{1/2},1/2}=1$. So the direct transition probability, the first term in Eq.~\eqref{tprob1}, can be obtained in the $m_J$ basis as
\begin{equation}
\abs{\bra{2,m_F} \hat U \ket{1,m_F}}^2
= A_{m_F}\abs{1-{\bra{5S_{1/2},-1/2} \hat{U}\ket{5S_{1/2},-1/2}} }^2,
\end{equation}
where the Clebsch-Gordon relation 
$C^{\frac{1}{2},\frac{3}{2},2}_{-\frac{1}{2},m_F+\frac{1}{2}}C^{\frac{1}{2},\frac{3}{2},1}_{-\frac{1}{2},m_F+\frac{1}{2}}=-C^{\frac{1}{2},\frac{3}{2},2}_{\frac{1}{2},m_F-\frac{1}{2}}C^{\frac{1}{2},\frac{3}{2},1}_{\frac{1}{2},m_F-\frac{1}{2}}$
is used to get
\begin{equation}
A_{m_F} = \abs{C^{\frac{1}{2},\frac{3}{2},2}_{\frac{1}{2},m_F- \frac{1}{2}} C^{\frac{1}{2},\frac{3}{2},1}_{\frac{1}{2},m_F- \frac{1}{2}} }^2=\abs{C^{\frac{1}{2},\frac{3}{2},2}_{-\frac{1}{2},m_F+ \frac{1}{2}} C^{\frac{1}{2},\frac{3}{2},1}_{-\frac{1}{2},m_F+ \frac{1}{2}} }^2.
\end{equation}

Similarly, the second term in Eq.~\eqref{tprob1} (the spontaneous emission) is obtained with the selection rule $\bra{5P_{1/2},m_J'}\hat U \ket{5S_{1/2},m_J}=\bra{5P_{1/2},1/2}\hat U \ket{5S_{1/2},-1/2}\delta_{m_J',1/2}\delta_{m_J,-1/2}$ to get
\begin{align}
&\bra{5P_{1/2},m_J'}\bra{I=3/2,m_I'}\hat U\ket{1,m_F}\nonumber\\
&~~~~~=\sum_{\substack{m_J=\pm1/2}}C^{\frac{1}{2},\frac{3}{2},1}_{m_J,m_F-m_J}\bra{5P_{1/2},m_J'}\hat U \ket{5S_{1/2},m_J}\langle{I=3/2,m_I'}|{I=3/2,m_F-m_J}\rangle\nonumber\\
&~~~~~=\sum_{\substack{m_J=\pm1/2}}C^{\frac{1}{2},\frac{3}{2},1}_{m_J,m_F-m_J}\bra{5P_{1/2},1/2}\hat U \ket{5S_{1/2},-1/2}\delta_{m_J',1/2}\delta_{m_J,-1/2}\langle{I=3/2,m_I'}|{I=3/2,m_F-m_J}\rangle
\nonumber\\
&~~~~~=\bra{5P_{1/2},1/2}\bra{I=3/2,m_F+1/2}\hat U\ket{1,m_F}\delta_{m_J',1/2}\delta_{m_I',m_F+1/2},
\end{align}
and it is further simplified using Eq.~\eqref{rel1} as
\begin{align}\label{rep}
&P_{\rm se}(1/2;\sigma^+,m_F)\abs{\bra{5P_{1/2},1/2}\bra{I=3/2,m_F+1/2}\hat U\ket{1,m_F}}^2\nonumber\\
&~~~~~=P_{\rm se}(1/2;\sigma^+,m_F)C^{\frac{1}{2},\frac{3}{2},1}_{-\frac{1}{2},m_F+\frac{1}{2}}\abs{ \bra{5P_{1/2}, 1/2} \hat{U} \ket{5S_{1/2}, -1/2}}^2.
\end{align}
$P_{\rm se}(1/2;\sigma^+,m_F)$ in Eq.~\eqref{rep} is the conditional probability of the spontaneous emission (from $\ket{5P_{1/2},1/2}$ to all ground $F=2$ sublevels) if the initial state is $\ket{1,m_F}$. Since the excited fine-structure state, $\ket{5P_{1/2}, 1/2}\ket{I=3/2,m_F+1/2}=\sum_{F'=1,2}C^{\frac{1}{2},\frac{3}{2},F'}_{\frac{1}{2},m_F+ \frac{1}{2}}\ket{F',m_F+1}$, decays to $F=1$ or $F=2$ with the rate proportional to the square of the transition dipole moment $\abs{\bra{F,m_F}er_q\ket{F',m_F'}}^2$, the conditional probability of the spontaneous emission to $F=2$ is given by
\begin{equation}
P_{\rm se}(1/2;\sigma^+,m_F)=\frac{\sum\limits_{\substack{F'=1,2,\\q=0, \pm1}}\abs{C^{\frac{1}{2},\frac{3}{2},F'}_{\frac{1}{2},m_F+ \frac{1}{2}}}^2\abs{\bra{2,m_F+1+q}er_q\ket{F',m_F+1}}^2}{\sum\limits_{\substack{F=1,2,\\ F'=1,2,\\ q=0,\pm1}}\abs{C^{\frac{1}{2},\frac{3}{2},F'}_{\frac{1}{2},m_F+ \frac{1}{2}}}^2\abs{\bra{F,m_F+1+q}er_q\ket{F',m_F+1}}^2}=\sum_{F'=1,2,q=0,\pm 1}  \abs{C^{\frac{1}{2},\frac{3}{2},F'}_{\frac{1}{2},m_F+\frac{1}{2}} D^{F',1,2}_{m_F+1,q}}^2,
\end{equation}
where the denominator is the sum of the decays to $F=1$ and $F=2$ (via $F'=1,2$), while the numerator is only to $F=2$. The transition dipole moment is defined between $J=1/2$ and $J'=1/2$ levels by
\begin{equation}
\bra{F,m_F}er_q\ket{F',m_F'}=\langle J\| er \| J'\rangle (-1)^{F'+\frac{1}{2}+1+I}D^{F',1,F}_{m_F',q}
\end{equation}
where $D^{j_1,j_2,j_3}_{m_1,m_2}$ = $\sqrt{2(2j_1+1)}\left\{\begin{matrix} 1/2 & 1/2 & 1 \\ j_1 & j_3 & 3/2\end{matrix}\right\}C^{j_1,j_2,j_3}_{m_1,m_2}$ with Wigner 6-j symbol expressed by curly brackets, and $q$ is the polarization index of $0$, $\pm1$ for $\pi$, $\sigma^\pm$, respectively.
Therefore, $B_{m_F}$, defined in \eqref{a_eq1}, is obtained
as 
\begin{equation}
B_{m_F} = \sum_{F'=1,2,q=0,\pm 1}  \abs{C^{\frac{1}{2},\frac{3}{2},F'}_{\frac{1}{2},m_F+\frac{1}{2}} D^{F',1,2}_{m_F+1,q}}^2 \abs{ C^{\frac{1}{2},\frac{3}{2},1}_{-\frac{1}{2},m_F+\frac{1}{2}} }^2.
\end{equation}

{\it The $\pi$ polarization case :} The transition probability for the $\pi$ polarization is given by
\begin{align}\label{tprob2}
&P(\mathcal{A},\Delta;\pi,m_F)\nonumber\\
&=\sum_{m_F''={\pm2}\pm1,0}\abs{\bra{2,m_F''} \hat U \ket{1,m_F}}^2 + \sum_{\substack{m_J'=\pm1/2,\\m_I'={ \pm3/2},\pm1/2}}P_{\rm se}(m_J';m_F,\pi)\abs{\bra{5P_{1/2},m_J'}\bra{I=3/2,m_I'}\hat U\ket{1,m_F}}^2.
\end{align}
The first term (the direct transition) vanishes as
\begin{align}
&\bra{2,m_F''} \hat U(\mathcal{A},\Delta;\pi) \ket{1,m_F}=\nonumber\\
&~~C^{\frac{1}{2},\frac{3}{2},2}_{\frac{1}{2},m_F-\frac{1}{2}}C^{\frac{1}{2},\frac{3}{2},1}_{\frac{1}{2},m_F-\frac{1}{2}}\bra{5S_{1/2},1/2} \hat{U}\ket{5S_{1/2},1/2}
+C^{\frac{1}{2},\frac{3}{2},2}_{-\frac{1}{2},m_F+\frac{1}{2}}C^{\frac{1}{2},\frac{3}{2},1}_{-\frac{1}{2},m_F+\frac{1}{2}}\bra{5S_{1/2},-1/2} \hat{U}\ket{5S_{1/2},-1/2}=0,
\end{align}
due to the facts $\bra{5S_{1/2},\pm1/2}\hat U \ket{5S_{1/2},\mp1/2}=0$ (selection rules) and $\bra{5S_{1/2},1/2}\hat U \ket{5S_{1/2},1/2}=\bra{5S_{1/2},-1/2}\hat U \ket{5S_{1/2},-1/ 2}$ (the same Rabi oscillation) with the Clebsch-Gordan relation. Likewise, using the selection rules and the relation
$\abs{\bra{5P_{1/2},-1/2} \hat U \ket{5S_{1/2},-1/2}}=\abs{\bra{5P_{1/2},1/2}\hat U \ket{5S_{1/2},1/2}}$, Eq.~\eqref{tprob2} becomes Eq.~\eqref{a_eq2}. The remaining procedure to obtain $E_{m_F}$ is similar to that for $B_{m_F}$, and finally we get
\begin{equation}
E_{m_F}=\sum_{\substack{F'=1,2, q=0,\pm 1,\\ m_J=\pm\frac{1}{2}}}  \abs{C^{\frac{1}{2},\frac{3}{2},F'}_{m_J,m_F-m_J} D^{F',1,2}_{m_F,q}}^2 \abs{ C^{\frac{1}{2},\frac{3}{2},1}_{m_J,m_F-m_J} }^2.
\end{equation}

\end{widetext}

\end{document}